# Real-Time Cure Monitoring via Carbon Nanotube Networks Enables Mechanical Property Optimization in Post-Cured Epoxy Resins

M. Jafarypouria[*], S.G. Abaimov

*Skolkovo Institute of Science and Technology*, *Moscow*, *Russia*

[*]Corresponding author: Milad.Jafarypouria@skoltech.ru

**Abstract**

This research presents a single-walled carbon nanotube (SWCNT)-enabled real-time monitoring system to optimize post-curing conditions (temperature and duration) for epoxy resin. This method can serve as an alternative to traditional methods like Differential Scanning Calorimetry (DSC), which is effective in measuring the degree of cure in polymers during industrial curing (manufacturer-recommended cure cycle). Two different programs using SWCNTs were employed to design the cure cycles for investigating the development of mechanical properties: Program A as the comparison of effects of varied duration of high-temperature curing and Program B as high-temperature curing followed by the varied duration of low-temperature post-curing. By correlating variation in the electrical resistance of SWCNT with curing stages, we illustrate that extending post-curing at 100°C for 24 hours after an initial 3-hour cure at 130°C increases tensile strength by 60% and elongation by 164% compared to industry standards. This approach not only improves mechanical performance but also enables precise, non-destructive cure-state detection, offering a scalable solution for high-performance composites in the aerospace and automotive sectors.

## 1. Introduction

Epoxy resins have been the most widely used thermosetting polymers since their early appearance in the 1930s. Epoxy resins represent a wide range of inherent properties, resulting from their highly reactive epoxy groups located in the terminal chains [1,2]. These thermosets' exceptional qualities make them appropriate for high-performance applications, such as those in the automobile and aircraft sectors. Nevertheless, because of the high cross-linking density that forms during curing, thermosetting polymer materials are often recognized for their inherent brittleness [3,4]. Therefore, the final properties of a structural epoxy system are not only highly influenced by the type and chemical structure of the monomers and the curing agent, which is the cross-linking precursor, but also by the curing conditions and external factors, such as the curing temperature, pressure, and so forth [5,6].

Selecting the appropriate cross-linking agent is a key strategy in the development of epoxy-type thermosetting polymers. Various options, such as amine hardeners, anhydride hardeners, and acid hardeners, have been explored and tested for this purpose [7–10]. Amine hardeners are the most used agent as they can provide cross-linking at relatively low temperatures [11,12]. It is important to carefully select the stoichiometric ratio between the hardener and epoxy resin. Any deviation from this ratio could result in an excess or deficiency of amine or epoxide groups, directly altering the characteristics of the epoxy system [13]. The glass transition temperature ($T_g$) of the resulting epoxy resins is especially affected by this phenomenon [14,15]. Numerous studies in this area have discovered that the stoichiometric point is when the ideal characteristics are reached [13,16].

| | |
|---|---|
| *List of symbols* | |
| $T_g$ | Glass transition temperature |
| TTT | Time-temperature-transformation diagram |
| $T_{cure}$ | Curing at various temperatures |
| $T_{g\infty}$ | Glass transition temperature of a fully reacted material |
| DSC | Differential scanning calorimetry |
| DEA | Dielectric analysis |
| FTIR | Fourier transform infrared radiation |
| SWCNT | Single-walled carbon nanotube |
| DC | Direct current |

Glass transition temperature ($T_g$) represents the temperature at which amorphous solids change from a glassy to a rubbery state. It reflects molecular mobility and depends on its degree of cure. In fact, the $T_g$ is a temperature range over which the mobility of the polymer chains increases significantly, causing the bulk material to transit from a glassy to a rubbery state. The temperature range over which this transition occurs varies greatly depending on the type of resin. Factors that influence $T_g$ include the composition of the resin molecule, crosslink density, polarity and molecular weight of the resin molecule, curing agent or catalyst, curing time, and curing temperature.

The relationship between temperature, time, and the transformation of a polymer during a curing process is typically depicted using an isothermal time-temperature-transformation (TTT) cure diagram [17,18]. The TTT diagram illustrates how the material transitions from a liquid or semi-liquid state to a solid state as it undergoes curing at a specific temperature over a defined period of time. The key characteristics of this diagram can be determined by measuring the event times that occur during isothermal curing at various temperatures, $T_{cure}$. These events are phase separation, gelation, vitrification, full cure, and devitrification. Achieving full cure status is most easily done by reacting above $T_{g\infty}$, and more slowly by curing below $T_{g\infty}$ to the full-cure state.

As the temperature at which curing occurs increases, $T_g$ of the network also increases steadily until it reaches $T_{g\infty}$. This is attributed to an increase in density (or decrease in free volume) and a predominance of short-range structural motions, which causes enhancement of the mechanical properties of the resin [7,19,20]. Once the temperature surpasses $T_{g\infty}$, the network will persist in a rubbery state following gelation, potentially leading to thermal degradation or oxidative crosslinking, resulting in degradation of the mechanical properties [21–24].

Enhancing the ultimate properties of a specific epoxy system can also be obtained by optimizing the curing process [25]. Specifically, employing lower temperatures during curing can lead to a thermosetting resin with a reduced $T_g$, as some reactive groups from either the epoxy resins or hardeners may not fully react [26]. Subsequently, post-curing is often necessary to maximize the final $T_g$ value [27]. Post-curing typically involves a higher temperature than that used during initial curing to achieve optimal cross-linking of the thermoset [28,29]. This procedure results in a resin with enhanced mechanical properties and, when combined with minimal shrinkage, provides superior stability. Moreover, post-curing is anticipated to increase the glass transition temperature, reduce residual stress, and decrease the tendency of outgassing [30].

Gupta et al. [27] illustrated that post-curing dramatically increases crosslink density in excess epoxy resins. Post-curing also reduces density [14,27]. Kong [14] found a decrease of 0.06 g/cm3 in a TGDDM/DDS-based formulation. This was attributed to evaporation of low-molecular-weight contaminants including water. Gupta et al. [27] found density decreased on the order of 0.01 g/cm3 over the entire formulation range.

The curing degree of resins can be evaluated using various conventional analytical techniques, including Fourier transform infrared (FTIR) spectroscopy [31], rotating or oscillating rheometry, and dielectric analysis (DEA) [32]. Among these, differential scanning calorimetry (DSC) is the most commonly used method. DSC is a thermal analysis technique that assesses crystallization, melting behavior, and other thermal transitions occurring in a sample when subjected to heating, cooling, or isothermal conditions [33]. This technique characterizes several properties of the resin, including the curing degree [34]. However, these methods often are destructive, typically limited to small sample sizes or weights, and may yield measurements that are not very precise [35,36]. Besides, they lack sensitivity when it comes to post-curing, where distinguishing between high and very high degrees of cure is crucial. Mahato et al. [37] illustrated using the DSC method that the epoxy resin they studied reaches nearly complete curing in about 3 hours at 130°C.

There is an increasing interest in utilizing carbon nanotubes (CNTs) as a versatile additive for sensing applications [38,39], driven by their remarkable electrical and thermal properties [40–46]. They can detect changes caused by resin infiltration [47,48] and polymer shrinkage during curing [48], as well as help assess the degree of polymerization [37,49–51] and the composition of the polymer [52–54]. In relation to the presented study, sensors on the base of SWCNTs demonstrate high sensitivity to the degree of cure at post-curing. They illustrated that significant post-curing processes take place even after being cured for several days at elevated temperatures which prompted the next step of this research to investigate the influence of lengthy post-curing on resin's mechanical properties.

This study addresses a critical gap in industrial epoxy curing practices by introducing single-walled carbon nanotubes (SWCNTs) as sensors for real-time cure-state monitoring. We first demonstrate that conventional manufacturer-recommended curing cycles (e.g., 3 hours at 130°C) fail to achieve full crosslinking, as evidenced by dynamic electrical resistance measurements from SWCNT networks. A non-destructive framework for optimizing post-curing conditions has been established by linking the consolidation of resistance to the completion of the curing process. These findings are then used to develop two post-curing regimes: Program A, which involves prolonged durations at 120°C, and Program B, which involves sequential high/low-temperature curing. This study introduces a paradigm shift toward data-driven curing techniques for high-performance epoxy systems, bridging the gap between laboratory-scale development and industrial scalability.

## 2. Mechanism of electrical conductivity in CNT/polymer nanocomposites

The significant contrast in electrical conductivity between insulating polymer matrices and conductive carbon nanotubes (CNTs) exhibits a percolation-like behavior in polymer/CNT nanocomposites. This behavior manifests as a sharp, power-law surge in conductivity once CNT concentrations exceed a critical threshold [55,56]. Two synergistic mechanisms govern this phenomenon: (1) nanoscale electron hopping (quantum tunneling) between adjacent CNTs, and (2) microscale conductive network formation [44,46]. At low CNT loadings (<0.05 wt.%), where inter-nanotube distances exceed 10 nm, electron hopping dominates charge transport. As CNT content increases, reduced inter-tube spacing (<2–5 nm) enables direct electrical connectivity between adjacent CNTs, establishing percolating conductive pathways [44,57–59]. Beyond this threshold, bulk conductivity becomes governed by the macroscopic network rather than localized tunneling events (Fig. 1).

The literature indicates that the percolation threshold for single-walled carbon nanotubes in polymer matrices is quite low (approximately 0.01% to 0.05%, depending, of course, on CNT length) when well-dispersed [60,61]. In the present study, a SWCNT loading of 0.6 wt.%—far exceeding the percolation threshold—ensures robust network formation.

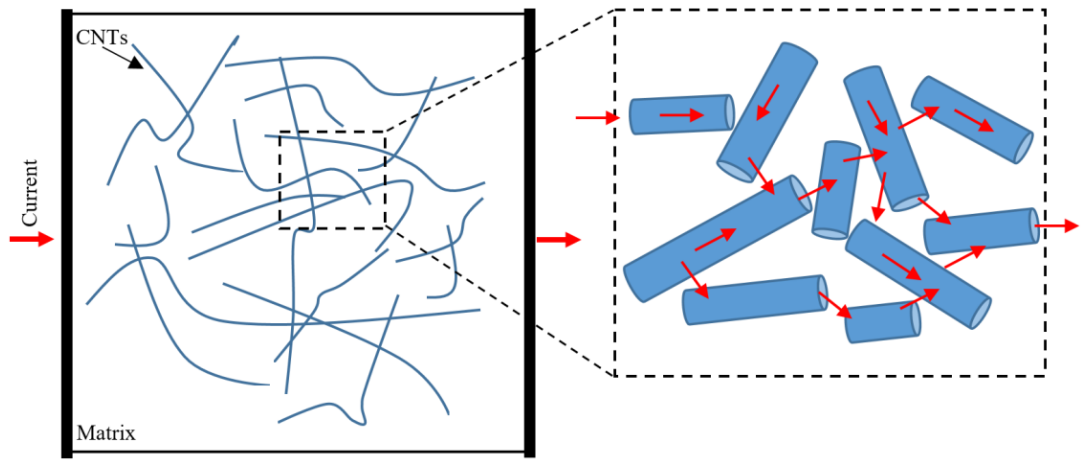

**Figure 1.** Schematic of electrical conduction pathway of CNT sensing network.

## 3. Curing State of the Epoxy Matrix Using SWCNTs

In this section, we present a method for monitoring the curing process by utilizing the electrical response (specifically, electrical resistance) of an epoxy resin filled with SWCNTs. Resistance measurements are carried out based on two different curing regimes: (i) constant temperature at 120°C for 72 hours, and (ii) varying temperature, spanning a range of 25-100°C, followed by the next 24 hours of post-curing and a new cycle of measurements, 8 cycles in total.

### 3.1 Materials and Methods

#### 3.1.1 Materials

The single-walled CNT masterbatch utilized in the experiment was TUBALL™ MATRIX 301 (OCSiAl), which is specifically formulated to enhance the electrical conductivity of epoxy, polyester, and polyurethane resins. The TUBALL™ SWCNT material is an effective filler to obtain conductive nanocomposites with good dispersion and low electrical resistance by melt-mixing [60]. For the fabrication of CNT/epoxy samples, we employed the epoxy resin system Biresin CR131, designed for high-performance fiber-reinforced polymer composite applications, as the matrix.

#### 3.1.2 Sample's Fabrication

The CNT masterbatch was combined with Biresin CR131 resin through shear mixing to achieve a target concentration of 0.6 wt.% CNTs at a room temperature of 25 °C and relative humidity of 30%. Three different stirring cycles, with varying speeds as detailed in Table 1, were employed. A low vacuum of 0.1 mbar was applied for 15 minutes between each stirring cycle to minimize air entrapment.

**Table 1.** Steps involved in the synthesis process for SWCNT/epoxy samples.

| **Mixed materials** | **Cycle 1** | | **Cycle 2** | | **Cycle 3** | |
|---|---|---|---|---|---|---|
| | Low speed + heating (45 ℃) (20 min) | 15 min Vacuum | High speed + heating (45 ℃) (60 min) | 15 min Vacuum | Low speed Without heating (20 min) | 15 min Vacuum |

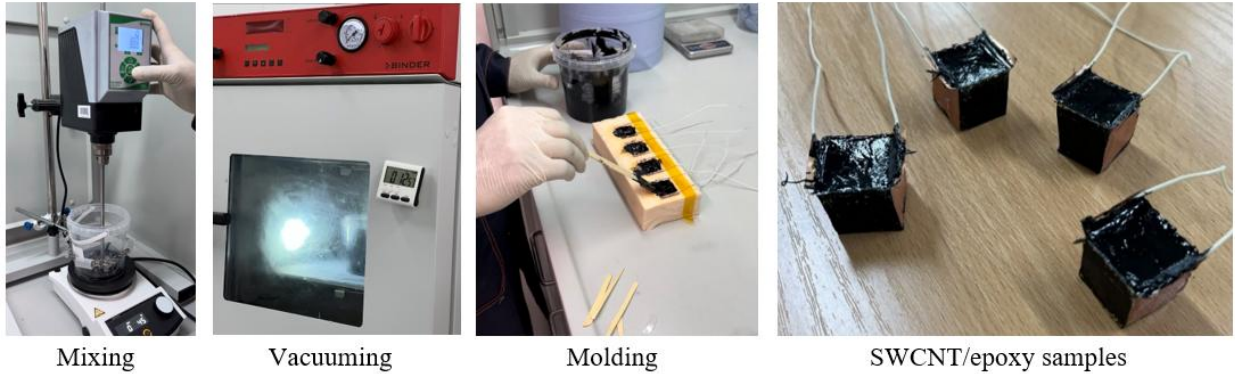

Mixing     Vacuuming     Molding     SWCNT/epoxy samples

**Figure 2.** Fabrication of SWCNT/epoxy samples.

In order to evaluate the electrical resistance of a sample, two copper tape electrodes were positioned on opposite sides of a cubic silicon mold 25 mm × 25 mm × 25 mm (see Figure 2). Five samples, for each curing regime, were molded and then cured at an industrial temperature of 130°C for 3 hours, as recommended by the resin manufacturer to achieve a high degree of cure.

To evaluate the quality of the nanofiller mixing, scanning electron microscopy was used to analyze the fracture surface of a sample (see Figure 3). Although minor agglomeration and bundling were observed, the nanofiller shows nearly perfect distribution and dispersion, exhibiting both isotropic and homogeneous characteristics. This is attributed to a high rotational speed of 3000 rpm led to both a more homogeneous SWCNT dispersion and a lower resistivity. Uçar et al. [60] showed that for a PMMA/SWCNT system, the number and size of agglomerations and the electrical resistivity decreased at higher rotation speeds.

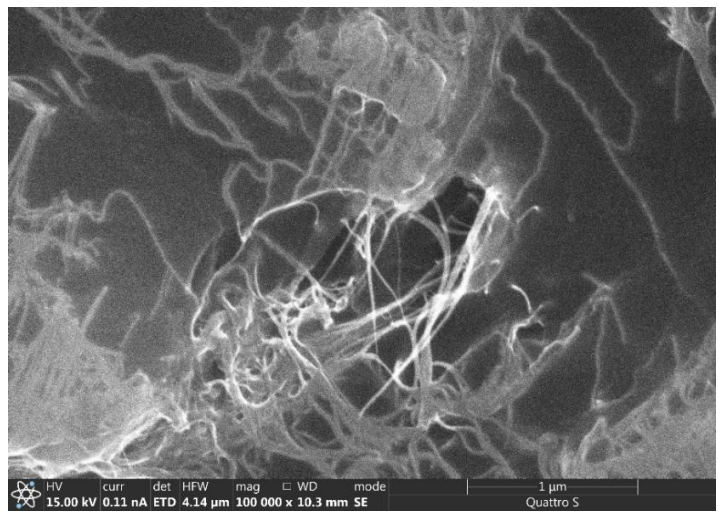

**Figure 3.** Scanning electron microscopy of a sample fracture surface.

*3.1.3 Electrical Resistance Measurements during Post-Curing*

DC electrical measurements were conducted on SWCNT/epoxy samples using a Keithley DMM6500, as illustrated in Figure 4. Before the post-curing heating, the initial resistances of the samples were recorded.

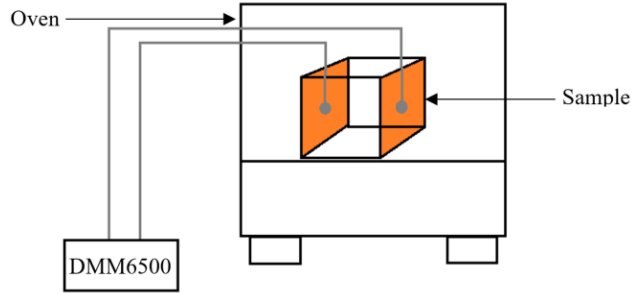

**Figure 4.** Schematic diagram of DC electrical measurements. Copper tape electrodes at the opposite faces of the sample to which conducting wires were soldered. The other end of conducting wires was connected to DMM 6500 for DC electrical measurement.

Procedure A: Five samples were post-cured for 72 hours straight at 120°C in a BINDER ED115. Every day, electrical resistance readings were taken while the samples were in the oven. Measurements took place for one hour every day, and reading changes were tracked during that time. This procedure refers to uninterrupted post-curing.

Procedure B: Five other samples were subjected to a temperature regime of 100°C for 8 days, during which the heating was temporarily interrupted for resistance measurements in the range of 25-100°C. To this purpose, the samples were slowly cooled to room temperature between each cycle. Electrical resistance was then measured at temperatures of 25°C (room temperature), 40°C, 60°C, 80°C, and 100°C, with each temperature maintained for 1 hour to ensure thermal equilibrium within the samples. After completing the measurements, the samples were further post-cured at 100°C for the remaining part of the day, and the next day the cycle repeated itself. This procedure was followed within 8 cycles. This procedure refers to interrupted post-curing.

## 3.2 Results of Electrical Resistance Change during Post-Curing

In this section, we present the obtained results of the DC electrical resistance measurements conducted during the samples' post-curing according to Procedure A (uninterrupted post-curing) and Procedure B (interrupted post-curing). Changes in conductivity result from variations in the dispersion and interaction of SWCNTs inside the matrix as the epoxy cures. This sensitivity can be used to determine whether the resin is fully cured because electrical properties (such as resistance in this study) will stabilize once curing is complete.

*3.2.1 Uninterrupted Post-Curing*

Figure 5 shows the measured electrical resistances for the five samples that were post-cured continuously at 120 °C for 72 hours. The resistance variation of a sample over an hour is represented by each connected set of markers. A clear decreasing trend is observed, indicating that the curing process is ongoing. We argue that the manufacturing method employed (like curing at 130°C for 3 hours) does not allow the samples to achieve a sufficiently high degree of cure, as evidenced by the noticeable changes in material properties during post-curing. It is evident that the curves only stabilize at 120°C for 72 hours, signifying the completion of the curing process.

All dependencies in Figure 5 display similar trends, but vary due to the scatter in the conductance values, which remain unchanged even after 72 hours of post-curing. This is attributed to scatter of differences in mixing, porosity, and the interface properties between the nanocomposite mixture and the copper electrode. These factors can both initiate gas bubble nucleation and affect CNT content due to the differences in surface energies between CNT-polymer and CNT-copper. Despite the observed scatter, the dependencies of post-curing time exhibit a distinct trend.

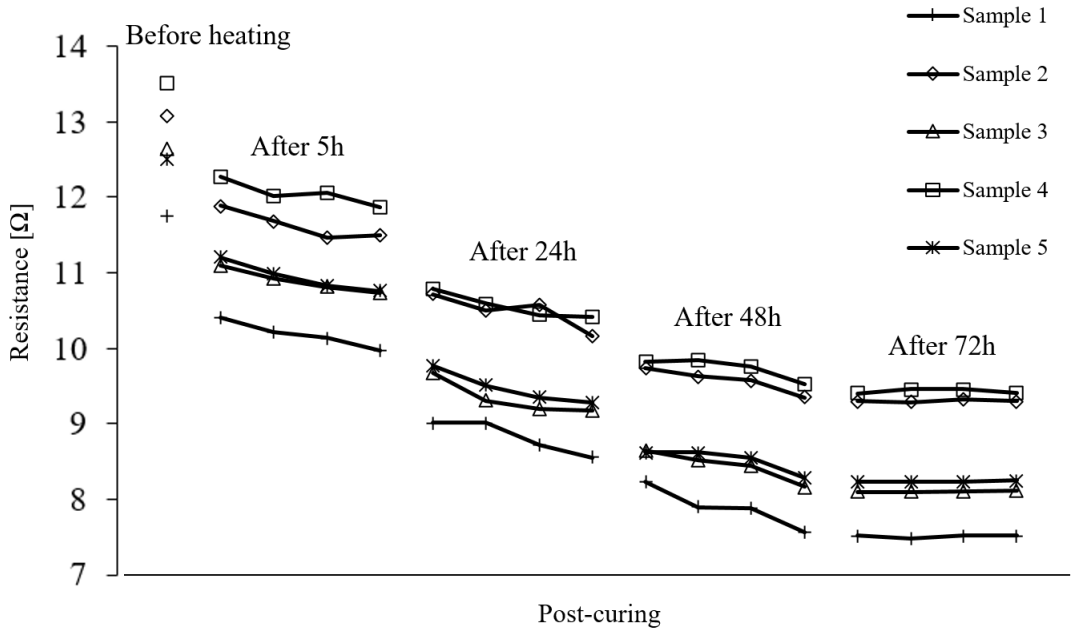

**Figure 5.** Change in electrical resistance of single-walled CNT/epoxy samples during the post-curing at T = 120°C for 72 h.

### 3.2.2 Interrupted Post-Curing

The dependence of the electrical resistance on temperature for the CNT/epoxy system was examined over eight thermal cycles, each involving 24 hours of post-curing at 100°C, interrupted for measurements. The experimental findings are illustrated in Figure 6. It is evident that the resistance decreases significantly with post-curing until the material achieves a fully cured steady state (as seen in cycles 7 and 8). Moreover, as it is seen, in cycle 1 the increase in resistance with temperature is non-monotonic; however, in the fully cured state, the increase in resistance with temperature becomes monotonic.

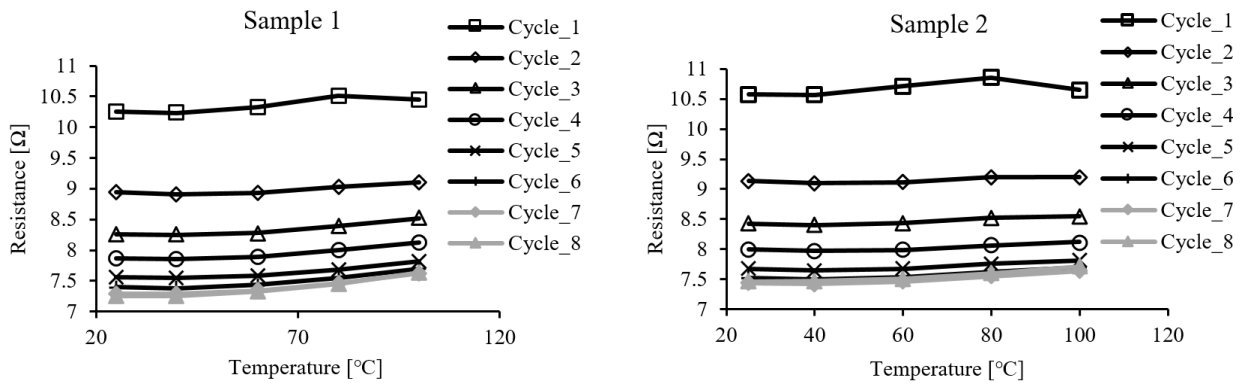

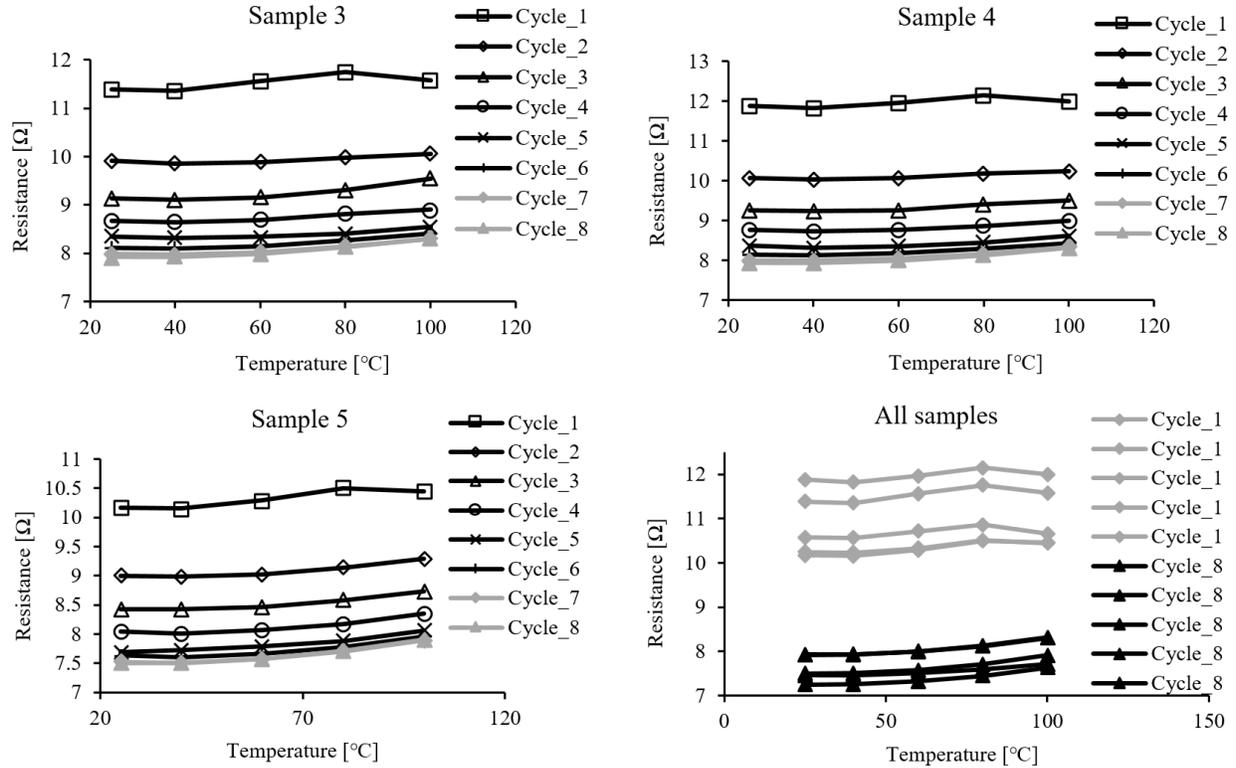

**Figure 6.** Change in electrical resistance with temperature at different post-curing cycles.

The significant decrease in resistance of SWCNT/epoxy samples during interrupted and uninterrupted post-curing until reaching a fully cured steady state can be attributed to several proposed mechanisms of cure status during a cure cycle [62–64]. Lee et al. [62] describe how the resistance change of a CNT network is related to the cure status, as the morphological change of CNT network plays a key role in cure status sensing.

- **Improved Polymer Network:** During post-curing, chemical processes happen in the epoxy resin. This increases the cross-linking of the polymer network. Consequently, there is less space between conducting channels, which facilitates electron transit.

- **Enhanced Electrical Conductivity:** Continuing the curing process causes chemical shrinkage, leading to reduced distance between CNTs. As a result, the resistance falls.

- **Reduction of Defects:** Defects/conductivity barriers can be decreased in the cured polymer using post-curing. This enhances the nanocomposite's charge transport properties by a more homogeneous structure.

- **Interface Optimization:** The interfacial properties between the SWCNTs and the epoxy matrix may be improved during post-curing, promoting electrical conduction.

## 4. Optimizing Mechanical Properties of Epoxy Matrix through Post-curing

This section evaluates the mechanical properties of the epoxy matrix through tensile and shear tests. Before performing mechanical testing, we ensure that the epoxy samples are completely cured using the post-curing conditions described in Section 3. The findings illustrate how the mechanical characteristics of the epoxy resin are improved by the optimized post-curing conditions.

### 4.1 Materials and Methods

*4.1.1 Materials*

Biresin CR131 epoxy resin and CH132-5 hardener, a two-part thermoset solution with thermal properties ($T_g$) up to 130°C, were used to fabricate the samples. This system is designed for high-performance fiber-reinforced polymer composite applications. Epoxy resin and hardener are both liquids with low viscosity.

*4.1.2 Samples' Fabrication*

The epoxy resin and hardener were added under the stoichiometric ratio 100:28 (*wt/wt*) following the manufacturer's recommendations. The resulting matrix was mixed well for up to ten minutes (stirring at a low speed of 300 rpm). The low vacuum of 0.1 bar was applied for 10 min to reduce air entrapment. Five different batches of specimens were then prepared using the silicon mold as shown in Figure 7. All samples were cured for 24 h at room temperature (25°C) and subsequently post-cured for different post-curing temperatures and durations according to Programs A and B listed in Table 2.

**Table 2.** Post-curing temperatures and durations for Program A and B following the optimized post-curing conditions presented in Section 3.

| Program | Post-curing duration | Post-curing temperature |
|---|---|---|
| A | 12 h | 120 °C |
|  | 1 day | 120 °C |
|  | 2 days | 120 °C |
|  | 3 days | 120 °C |
|  | 1 day | 130 °C |
| B | 3 h | 130 °C |
|  | 3 h, 5 h | 130 °C, 100 °C |
|  | 3 h, 1 day | 130 °C, 100 °C |
|  | 3 h, 3 days | 130 °C, 100 °C |
|  | 3 h, 7 days | 130 °C, 100 °C |

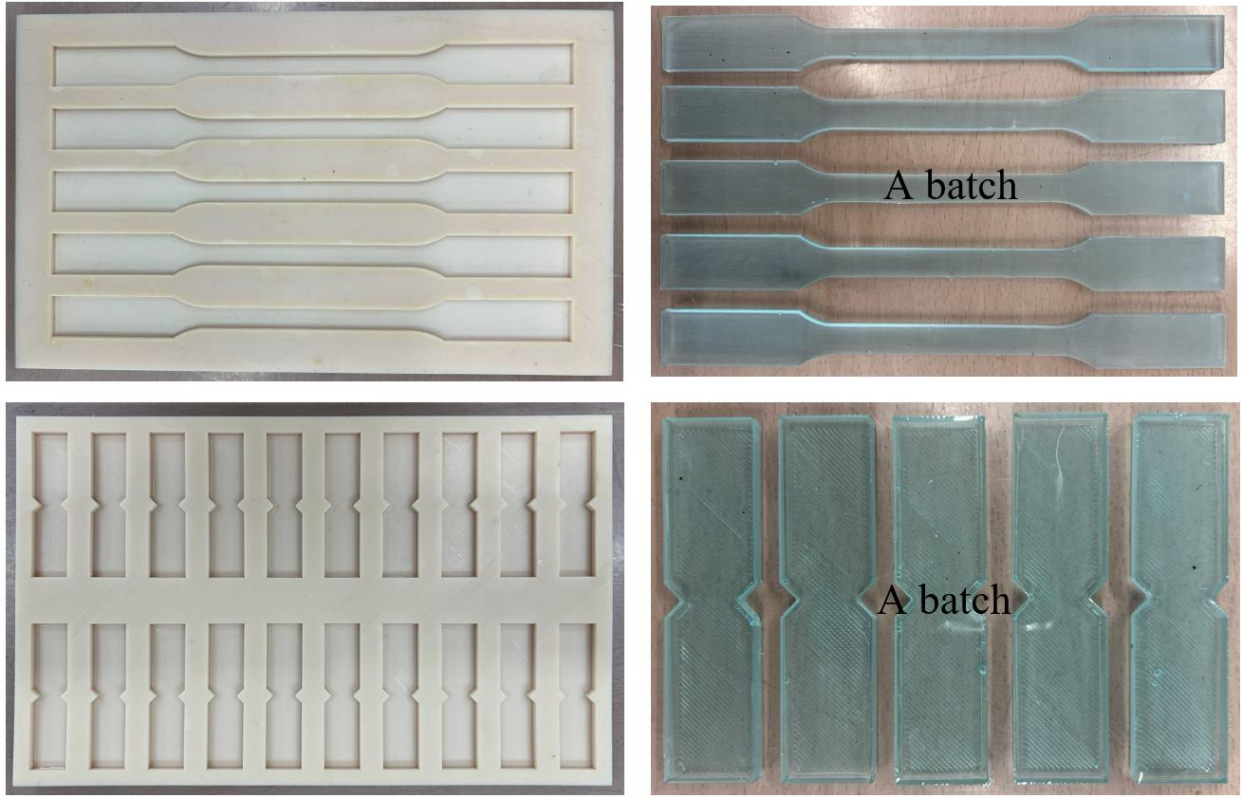

**Figure 7.** The silicon mold and test specimens. Five samples for each post-curing regime within both programs, referred to as a batch.

## 4.2 Mechanical Tests

The sample dimensions for tensile and shear tests follow international standard ISO 527-2 type 1A and ASTM D5379/D5379M-12, respectively (see Figure 8). The tests were performed according to standards on rectangular pieces in a mechanical universal testing machine INSTRON 5969 with a 50 kN load cell. The cross-head speed was set at 1 $mm.min^{-1}$ for tensile and shear tests. Figure 8 shows the equipment setup for tensile and shear tests.

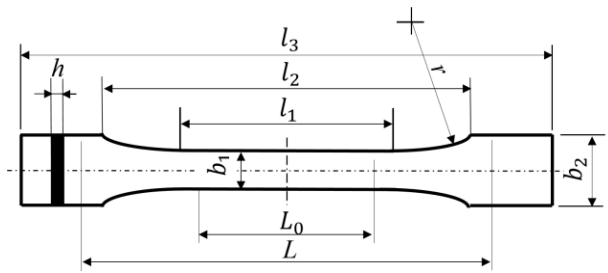

| | |
|---|---|
| Overall length ($l_3$) | 170 |
| Length of narrow parallel-sided portion ($l_1$) | 80 ± 2 |
| Radius ($r$) | 24 ± 1 |
| Distance between broad parallel-sided portions ($l_2$) | 109.3 ± 3.2 |
| Width at ends ($b_2$) | 20 ± 0.2 |
| Width at narrow portion ($b_1$) | 10 ± 0.2 |
| Preferred thickness ($h$) | 4 ± 0.2 |
| Gauge length ($l_0$) | 75 ± 0.5 |
| Initial distance between grips ($L$) | 115 ± 1 |

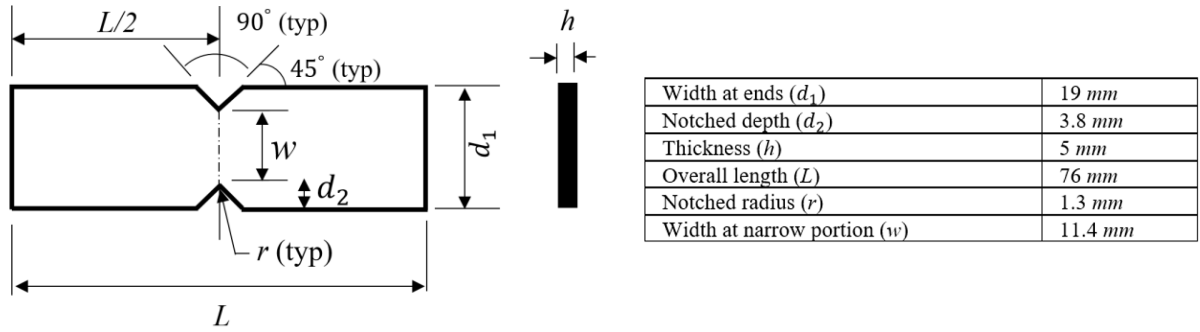

**Figure 8.** Dimensions of the specimens: a) tensile test standard ISO 527-2 type 1A, b) shear test standard ASTM D5379/D5379M-12.

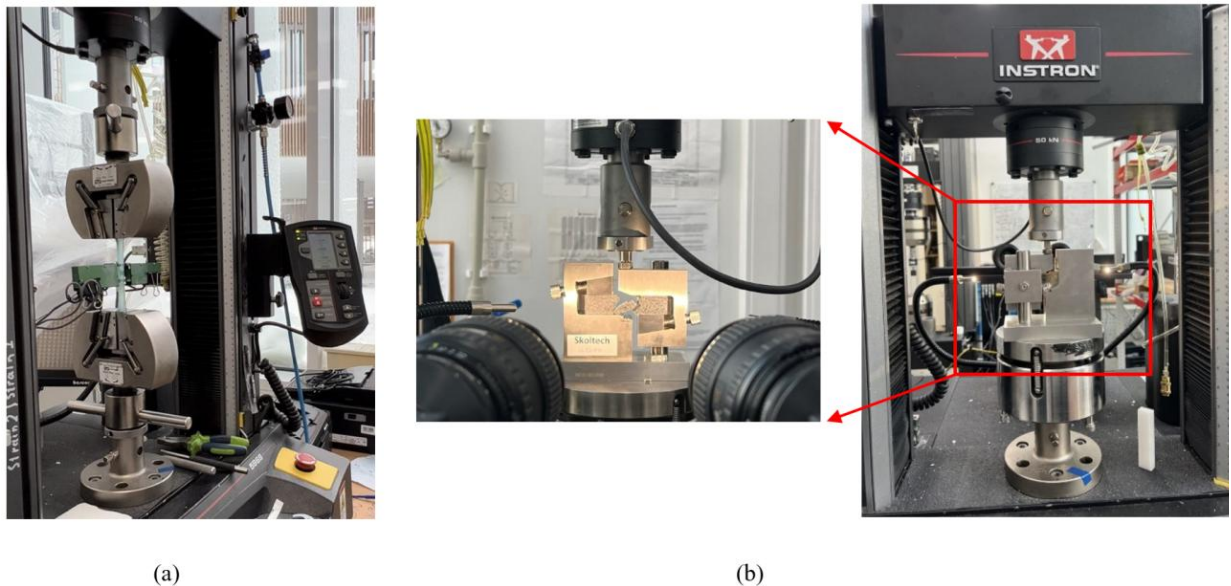

**Figure 9.** Equipment setup in the mechanical universal testing machine INSTRON 5969: a) tensile test, b) shear test.

### 4.3 Results of Mechanical Tests of Epoxy Matrix

In this section, we present the obtained results of the mechanical tensile test according to Programs A and B of post-cured samples. Subsequently, we will perform a mechanical shear test based on the optimal program choice between A and B. The mechanical properties (strength, Young's modulus, and Poisson ratio) were measured for each post-cured duration. Five specimens (a batch) were used for each case for statistical averaging of results.

*4.3.1 Program A: Tensile Test*

Figure 10a shows the stress vs. strain curves for the specimens with maximal ultimate stress in each batch that were heated based on Program A as a function of the cure temperature. The data obtained illustrate an increase in resin strength with increasing post-cure duration at temperature below $T_g$ (120°C).

As the cure temperature increases, equaling $T_g$ (130°C), the resin strength decreases significantly. This is attributed to thermal degradation or oxidative crosslinking that occurs once the temperature becomes equal to or surpasses $T_g$, resulting in the degradation of the mechanical properties. Figure 10b depicts the average stress at the break of five samples in each batch for Program A. As it is seen, post-curing at 120°C for 2 and 3 days increases the strength of the epoxy resin on average (by about 16.4%). However, a significant reduction is seen due to the material degradation after 1 day at 130°C (by about 33.6% compared to the maximum average stress observed on day 2 at 120 °C). Figure 10c displays the average elongation at break of five samples in each batch for Program A. Post-curing at 120°C for 2 and 3 days evidently increases the elongation of the epoxy resin on average (by about 10%). However, a substantial reduction is observed due to material degradation after 1 day at 130°C (approximately 55% compared to the maximum average stress observed on day 3 at 120°C).

Figure 11 represents the average Young's modulus ($E$) and Poisson ratio ($v$) of five samples in each batch for Program A. From Figure 11a, it is seen that Young's modulus on day 2 at 120°C shows an improvement of about 13.3% compared to 12 h post-curing at 120°C. An interesting point was observed for the temperature equal to $T_g$ (130°C), at which it demonstrates about a 33.6% increase of Young's modulus in comparison with 12 h post-curing at 120°C. This can be attributed to two factors: the first is a significant rise in crosslink density during post-curing, and the second is a drop in weight density caused by the evaporation of low-molecular-weight impurities such as water. As a result, extended curing times lead to material deterioration and increased brittleness with Young's modulus rise, while the resin's strength falls. Figure 11b illustrates that there is no pronounced difference in the Poisson ratio for different post-curing temperatures as well as durations.

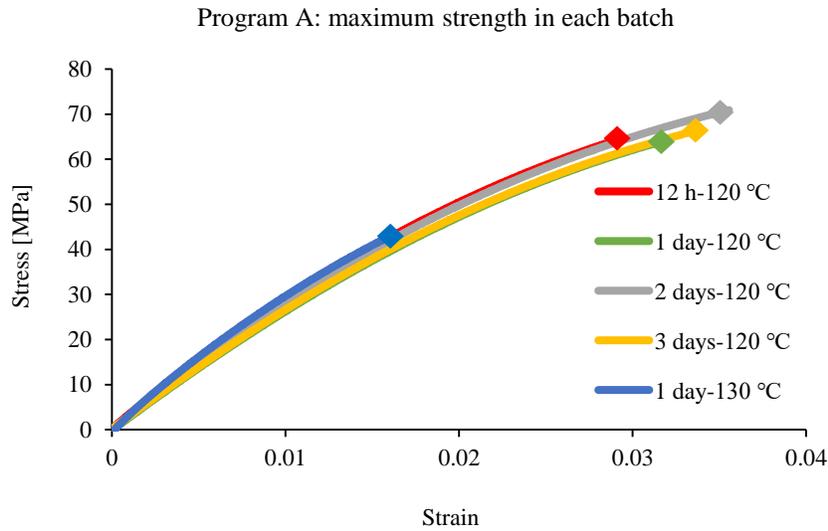

(a)

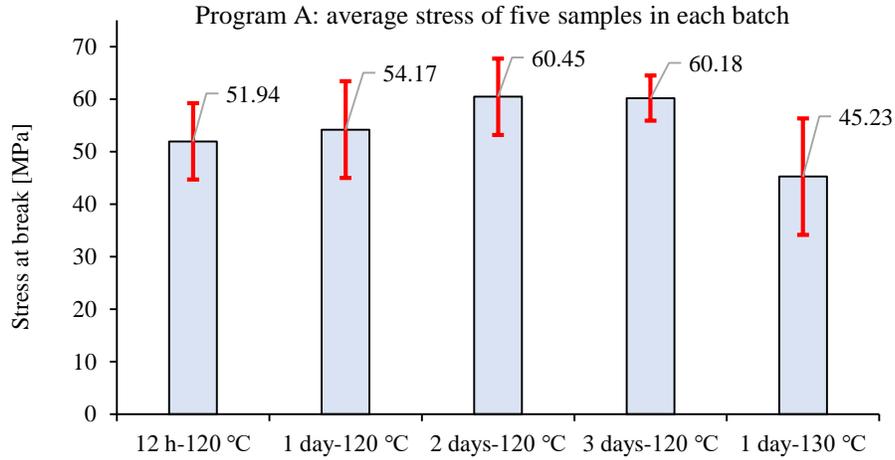

(b)

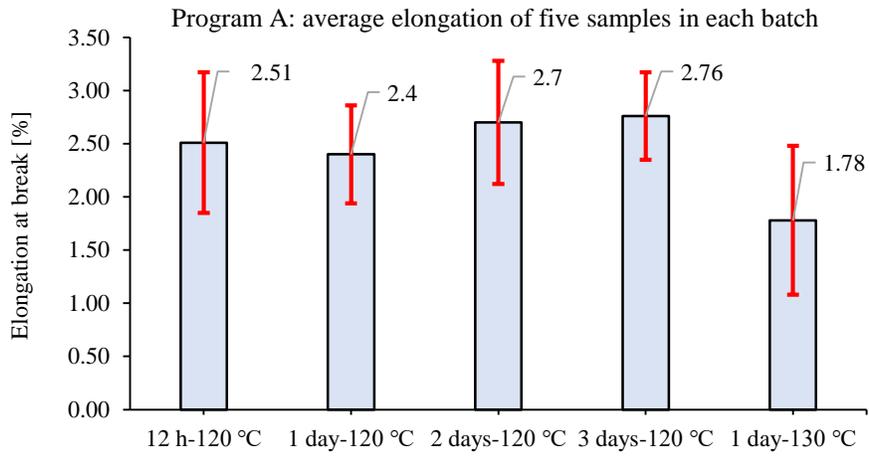

(c)

**Figure 10.** Tensile testing of epoxy resin CR131 samples for Program A; a) stress-strain curves with maximal ultimate stress in each batch, b) average stress of five samples in each batch at break, and c). average elongation of five samples in each batch at break.

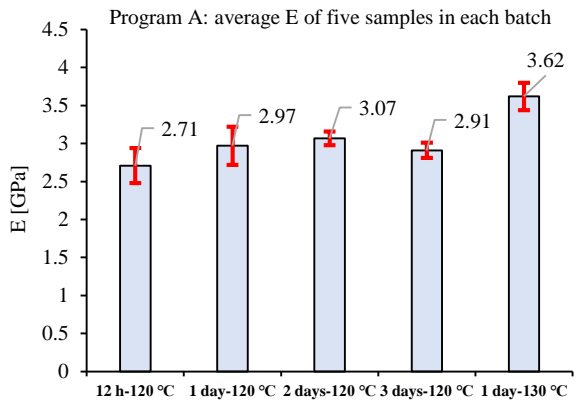

(a)

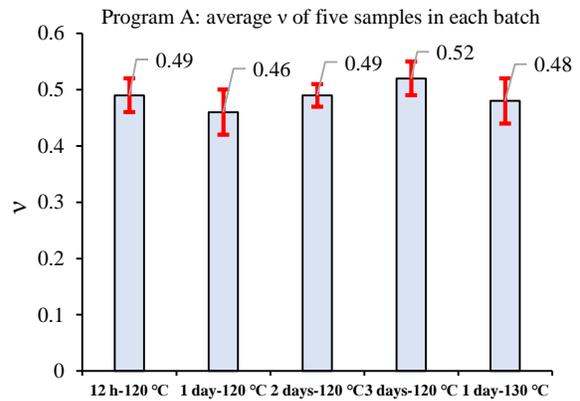

(b)

**Figure 11.** Tensile testing of epoxy resin CR131 samples for Program A; a) average Young's modulus of five samples in each batch, b) average Poisson's ratio of five samples in each batch.

*4.3.2 Program B: Tensile Test*

Figure 12a shows the stress vs. strain curves for the specimens with maximal ultimate stress in each batch that was heated based on program B as a function of the cure temperature. The results show an increase in resin strength with increasing post-cure duration to 1 day at 100°C after curing at 130°C for 3 h. However, as the post-cure duration is increased above 1 day, a clear degradation in the mechanical properties was observed (for 3 h at 130°C + 3 days at 100°C and 3 h at 130°C + 7 days at 100°C). When keeping an epoxy resin at an elevated temperature for too long, the strength is reduced because of the thermal degradation of the epoxy. Thermal degradation consists of the rupturing of chains, which can split off volatiles (weight loss) and/or generate reactive sites in the material that will cross-link again. After longer times, the weight loss becomes more pronounced [20], and the repeated cross-linking results in char formation, a high carbon content residue. The higher the curing temperature is, the faster this degradation process follows the full cure of the epoxy. Figure 12b shows the average stress at break of five samples in each batch for Program B. As it is seen, post-curing for 3 h at 130°C + 1 day at 100°C dramatically increases the strength of the epoxy resin on average (about 60%) compared to only 3 h at 130°C (industrial practice). However, a significant reduction was observed due to the material degradation for 3 h at 130°C + 7 days at 100°C as a result of keeping for too long at elevated temperature. Figure 12c displays the average elongation at the break of five samples in each batch for Program B. It is evident that post-curing for 3 hours at 130°C + 1 day at 100°C notably increases the elongation of the epoxy resin on average (by about 164%) compared to only 3 hours at 130°C (the industrial practice). However, a significant reduction is observed due to material degradation for 3 hours at 130°C + 7 days at 100°C, as a result of prolonged exposure to elevated temperature.

Figure 13 depicts the average Young's modulus ($E$) and Poisson's ratio ($v$) of five samples in each batch for Program B. It is seen that with longer post-curing durations both Young's modulus and Poisson ratio illustrate an important increase of up to 10% and 8% compared to the industrial practice (3 h at 130°C), respectively. As the material cures over an extended period, it undergoes degradation which leads to increased brittleness. Consequently, resin's strength diminishes while Young's modulus and Poisson's ratio rise.

Figure 14 compares the obtained stress vs. strain curves for maximal ultimate stress in a batch for two regimes, 2 days at 120°C in Program A and 3 h at 130°C + a day at 100°C in Program B, which demonstrated the best strengths in these programs. The results present that Program B exhibits approximately 17% higher ultimate elongation compared to Program A. Additionally, there is a minor 2% increase in stress at the break for Program B in comparison to Program A for these regimes. The findings suggest that post-curing epoxy resin CR131 for 3 h at 130°C + 1 day at 100°C as in high temperature curing followed by lengthy post-curing at low temperatures, demonstrates better mechanical properties based on our proposed post-curing programs.

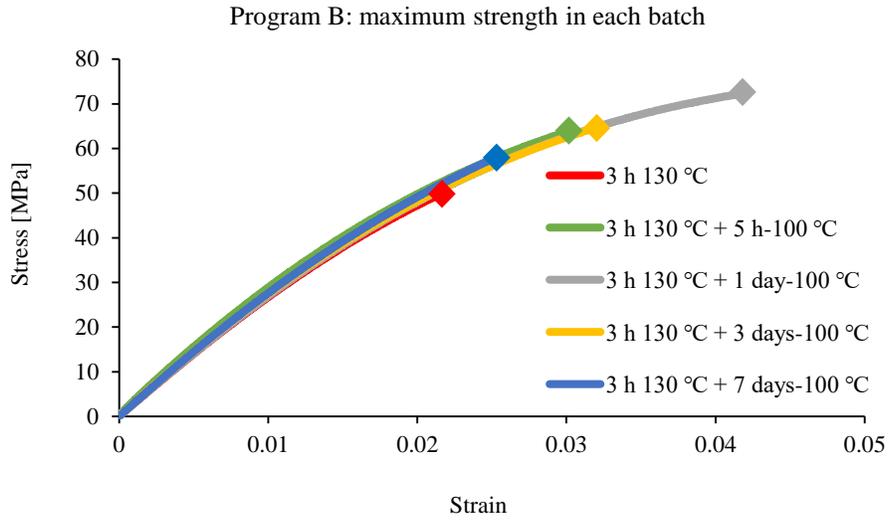

(a)

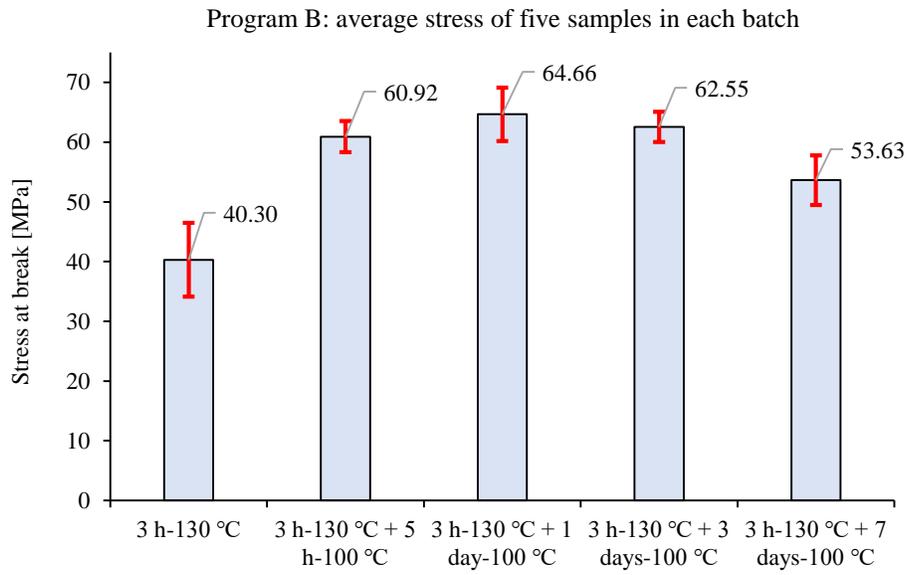

(b)

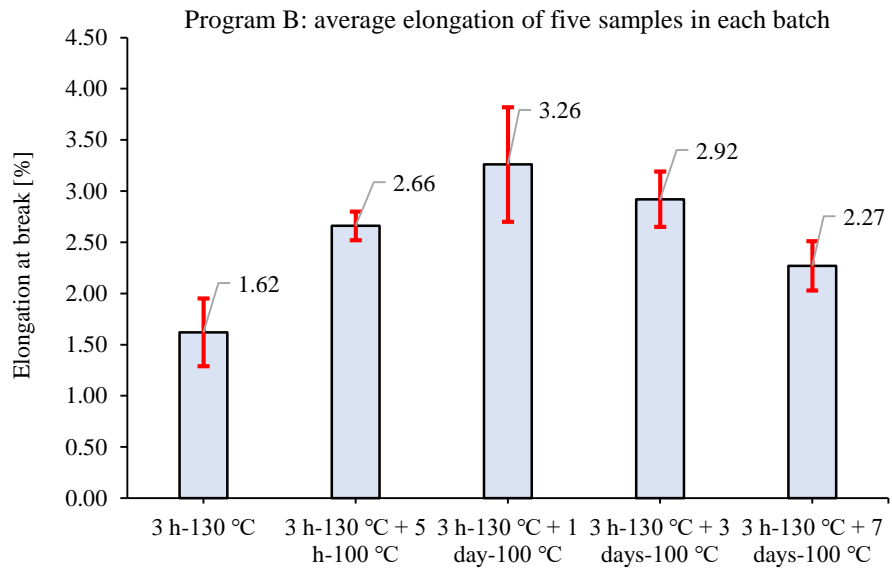

(c)

**Figure 12.** Tensile testing of epoxy resin CR131 samples for program B; a) stress-strain curves for maximal ultimate stress in each batch, b) average stress of five samples in each batch at break, and c) average elongation of five samples in each batch at break.

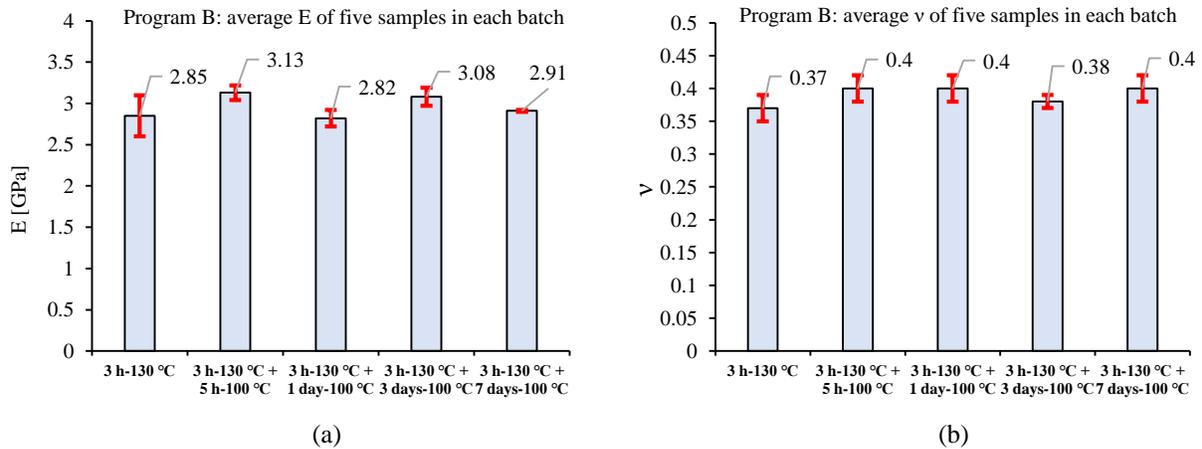

(a)          (b)

**Figure 13.** Tensile testing of epoxy resin CR131 samples for Program B; a) average Young's modulus of five samples in each batch, b) average Poisson's ratio of five samples in each batch.

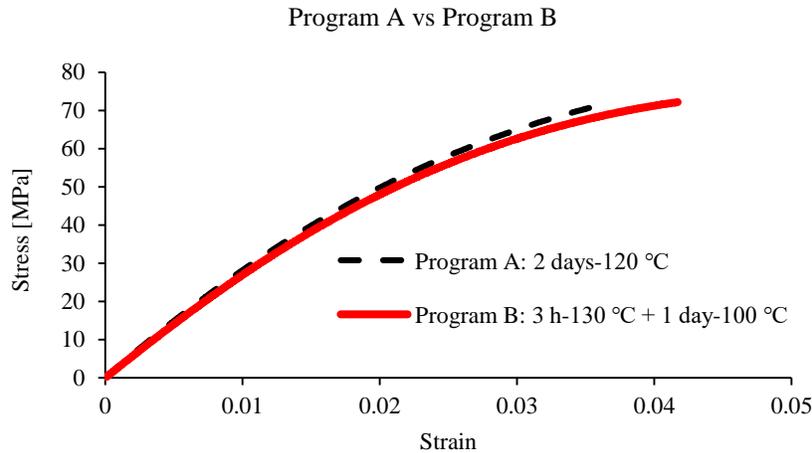

**Figure 14.** Tensile testing of epoxy resin CR131 samples: Program A vs Program B comparison based on the best observed post-curing regime in each program.

*4.3.3 Program B: Shear Test*

According to our observation in section 4.3.1 and section 4.3.2, a mechanical shear test was done based on program B as provided better results in tensile test.

Figure 15a depicts the shear stress vs. shear strain curves for the specimens with maximal ultimate shear stress in each batch that were heated based on Program B as a function of the cure temperature. The results show an increase in resin shear strength with increasing post-cure duration. However, as the post-cure duration is further increased, a clear degradation in the mechanical properties was observed. Again, this is due to keeping an epoxy resin at an elevated temperature for too long that leads to the material degradation. Figure 15b shows the average shear stress at break of five samples in each batch. The average results illustrate an increase of about 14% in resin shear strength with increasing post-cure duration to 1 day at 100°C after curing at 130°C for 3 h. Figure 15c illustrates the average shear strain at break of five samples in each batch. The average results indicate an increase of about 16% in resin shear strain with the extension of post-cure duration to 1 day at 100°C after curing at 130°C for 3 hours.

Figure 16 depicts average shear modulus (*G*) of five samples in each batch based on Program B. It is seen that with longer post-curing durations, the maximum relative difference (3 h at 130°C + 5 h at 100°C) shows an increase of about 6% compared to the industrial practice (3 h at 130°C). As the material cures over an extended period, it undergoes degradation which leads to increased brittleness, results in decreasing resin's strength while shear modulus ratio rise.

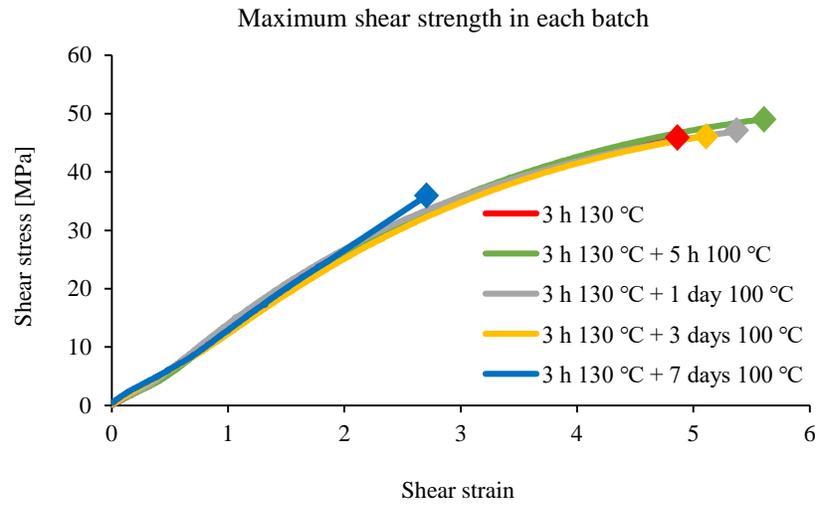

(a)

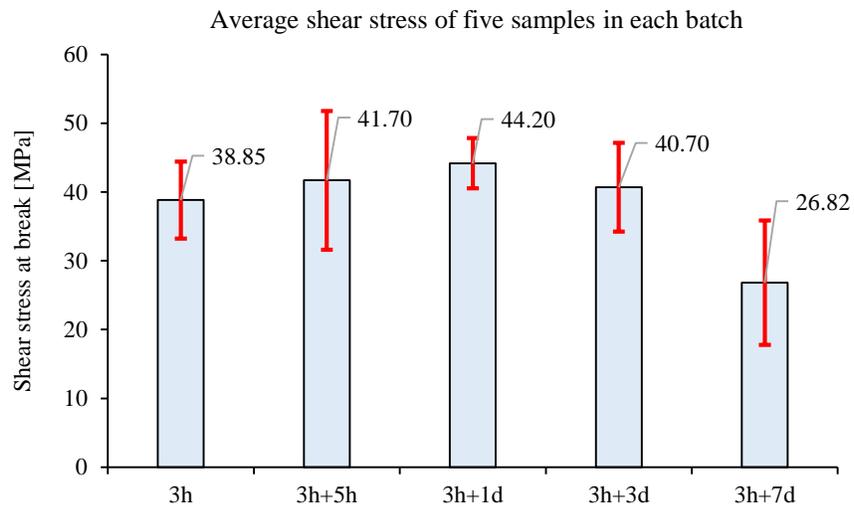

(b)

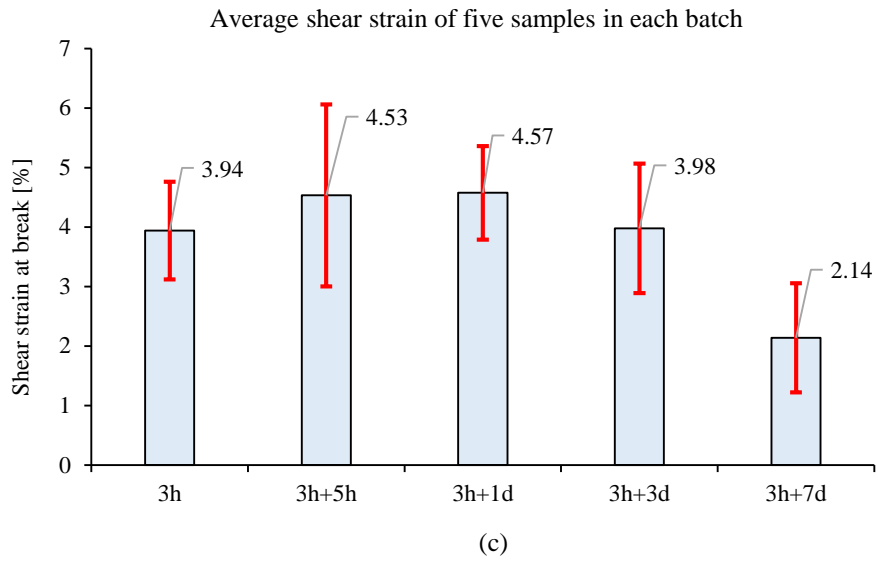

(c)

**Figure 15.** Shear testing of epoxy resin CR131 samples for program B; a) shear stress-shear strain curves for maximal ultimate shear stress in each batch, b) average shear stress of five samples in each batch at break, and c) average shear strain of five samples in each batch at break.

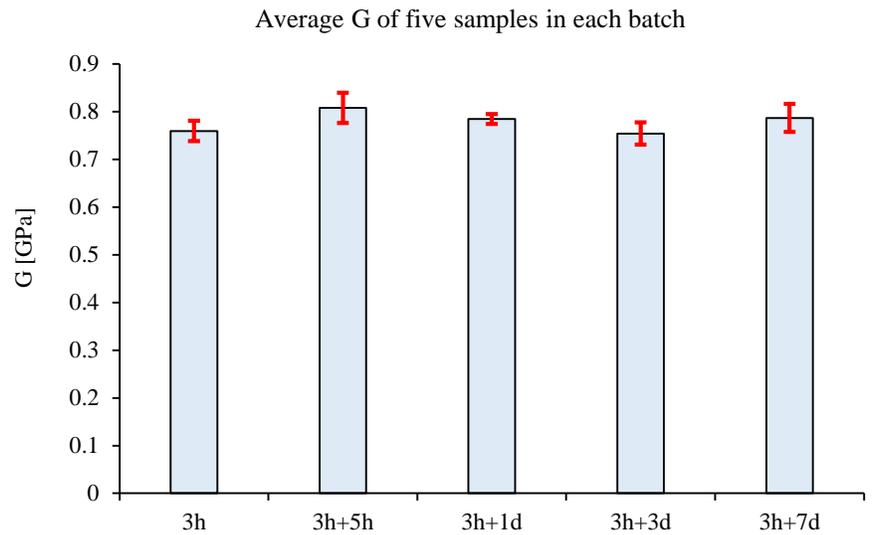

**Figure 16.** Shear testing of epoxy resin CR131 samples for program B; average shear modulus of five samples in each batch.

Table 3 summarizes the values of the mechanical properties of epoxy resin CR131 for both post-curing programs A and B, obtained from average values of five samples in each batch.

**Table 3.** Tensile strength, shear strength and elastic modulus for specimens tested at different post-curing conditions (average ± standard deviation).

| Program | Post-curing duration | Post-curing temperature | Tensile strength [MPa] | Shear strength [MPa] | Elongation [%] | stiffness $E$ [GPa] | Shear modulus $G$ [GPa] | Poisson ratio $\nu$ |
|---|---|---|---|---|---|---|---|---|
| A | 12 h | 120 °C | 51.94 ± 7.26 | - | 2.51 ± 0.66 | 2.71 ± 0.23 | - | 0.49 ± 0.03 |
|  | 1 day | 120 °C | 54.17 ± 9.2 | - | 2.4 ± 0.46 | 2.97 ± 0.25 | - | 0.46 ± 0.04 |
|  | 2 days | 120 °C | 60.45 ± 7.27 | - | 2.7 ± 0.58 | 3.07 ± 0.09 | - | 0.49 ± 0.02 |
|  | 3 days | 120 °C | 60.18 ± 4.28 | - | 2.76 ± 0.41 | 2.91 ± 0.1 | - | 0.52 ± 0.03 |
|  | 1 day | 130 °C | 45.23 ± 11.07 | - | 1.78 ± 0.70 | 3.62 ± 0.18 | - | 0.48 ± 0.04 |
| B | 3 h | 130 °C | 40.30 ± 6.19 | 38.85 ± 5.6 | 1.62 ± 0.33 | 2.85 ± 0.25 | 0.76 ± 0.02 | 0.37 ± 0.02 |
|  | 3 h, 5 h | 130 °C, 100 °C | 60.92 ± 2.61 | 41.7 ± 10.1 | 2.66 ± 0.14 | 3.13 ± 0.09 | 0.81 ± 0.03 | 0.4 ± 0.02 |
|  | 3 h, 1 day | 130 °C, 100 °C | 64.66 ± 4.5 | 44.2 ± 3.66 | 3.26 ± 0.56 | 2.82 ± 0.1 | 0.78 ± 0.01 | 0.4 ± 0.02 |
|  | 3 h, 3 days | 130 °C, 100 °C | 62.55 ± 2.54 | 40.7 ± 6.46 | 2.92 ± 0.27 | 3.08 ± 0.11 | 0.75 ± 0.02 | 0.38 ± 0.01 |
|  | 3 h, 7 days | 130 °C, 100 °C | 53.63 ± 4.14 | 26.82 ± 9.05 | 2.27 ± 0.24 | 2.91 ± 0.01 | 0.79 ± 0.03 | 0.4 ± 0.02 |

## 5. Conclusion

By integrating SWCNTs as real-time cure monitors, this work establishes a data-driven framework for optimizing the post-curing of the matrix and as a result the mechanical properties of the epoxy resin. While traditional means, like Differential Scanning Calorimetry (DSC), to measure the degree of cure of a polymer work well at industrial curing, they possess low sensitivity at post-curing, where high and very high degrees of cure need to be differentiated. The present method provides a scalable solution for high-performance composites by improving mechanical performance and enabling accurate, non-destructive cure-state verification.

In order to study the development of mechanical properties, two different programs were used to design the cure cycles. Program A with varying curing durations at a high-temperature of 120°C, while Program B involved post-curing at 100°C for 24 hours following an initial 3-hour cure at 130°C. The findings demonstrate that Program B improves tensile strength by 60% and elongation by 164% in comparison to industry standards by establishing a link between the electrical resistance variation of SWCNT and curing phases.

**References**

[1] Jin, F. L., Li, X., & Park, S. J. (2015). Synthesis and application of epoxy resins: A review. *Journal of industrial and engineering chemistry*, *29*, 1-11.

[2] Yu, S., Li, X., Guo, X., Li, Z., & Zou, M. (2019). Curing and characteristics of N, N, N′, N′-tetraepoxypropyl-4, 4′-diaminodiphenylmethane epoxy resin-based buoyancy material. *Polymers*, *11*(7), 1137.

[3] Jin, F. L., & Park, S. J. (2008). Thermomechanical behavior of epoxy resins modified with epoxidized vegetable oils. *Polymer International*, *57*(4), 577-583.

[4] Lascano, D., Quiles-Carrillo, L., Torres-Giner, S., Boronat, T., & Montanes, N. (2019). Optimization of the curing and post-curing conditions for the manufacturing of partially bio-based epoxy resins with improved toughness. *Polymers*, *11* (8), 1354.

[5] Kim, R. W., Kim, C. M., Hwang, K. H., & Kim, S. R. (2019). Embedded based real-time monitoring in the high-pressure resin transfer molding process for CFRP. *Applied Sciences*, *9* (9), 1795.

[6] Rudawska, A. (2019). The impact of the seasoning conditions on mechanical properties of modified and unmodified epoxy adhesive compounds. *Polymers*, *11*(5), 804.

[7] Enns, J. B., & Gillham, J. K. (1983). Effect of the extent of cure on the modulus, glass transition, water absorptio, and density of an amine-cured epoxy. *Journal of applied polymer science*, *28*(9), 2831-2846.

[8] Ivankovic, M., Incarnato, L., Kenny, J. M., & Nicolais, L. (2003). Curing kinetics and chemorheology of the epoxy/anhydride system. *Journal of applied polymer science*, *90* (11), 3012-3019.

[9] Zilg, C., Mülhaupt, R., & Finter, J. (1999). Morphology and toughness/stiffness balance of nanocomposites based upon anhydride-cured epoxy resins and layered silicates. *Macromolecular Chemistry and Physics*, *200*(3), 661-670.

[10] Zheng, T., Wang, X., Lu, C., Zhang, X., Ji, Y., Bai, C., ... & Qiao, Y. (2019). Studies on curing kinetics and tensile properties of silica-filled phenolic amine/epoxy resin nanocomposite. *Polymers*, *11* (4), 680.

[11] Guermazi, N., Haddar, N., Elleuch, K., & Ayedi, H. F. (2014). Investigations on the fabrication and the characterization of glass/epoxy, carbon/epoxy and hybrid composites used in the reinforcement and the repair of aeronautic structures. *Materials & Design (1980-2015)*, *56*, 714-724.

[12] Park, S. J., Seo, M. K., & Lee, J. R. (2000). Isothermal cure kinetics of epoxy/phenol-novolac resin blend system initiated by cationic latent thermal catalyst. *Journal of Polymer Science Part A: Polymer Chemistry*, *38*(16), 2945-2956.

[13] Mostovoy, S., & Ripling, E. J. (1966). Fracture toughness of an epoxy system. *Journal of Applied Polymer Science*, *10* (9), 1351-1371.

[14] Kong, E. S. W. (2005). Physical aging in epoxy matrices and composites. In *Epoxy resins and composites IV* (pp. 125-171). Berlin, Heidelberg: Springer Berlin Heidelberg.

[15] Fu, K., Xie, Q., Lü, F., Duan, Q., Wang, X., Zhu, Q., & Huang, Z. (2019). Molecular dynamics simulation and experimental studies on the thermomechanical properties of epoxy resin with different anhydride curing agents. *Polymers*, *11* (6), 975.

[16] Kenyon, A. S., & Nielsen, L. E. (1969). Characterization of network structure of epoxy resins by dynamic mechanical and liquid swelling tests. *Journal of Macromolecular Science—Chemistry*, *3*(2), 275-295.

[17] Carbas, R. J. C., Marques, E. A. S., Da Silva, L. F. M., & Lopes, A. M. (2014). Effect of cure temperature on the glass transition temperature and mechanical properties of epoxy adhesives. *The Journal of Adhesion*, *90*(1), 104-119.

[18] Gillham, J. K. (1986). Formation and properties of thermosetting and high Tg polymeric materials. *Polymer Engineering & Science*, *26* (20), 1429-1433.

[19] Varley, R. J., Hodgkin, J. H., & Simon, G. P. (2000). Toughening of trifunctional epoxy system. V. Structure–property relationships of neat resin. *Journal of applied polymer science*, *77*(2), 237-248.

[20] Ziaee, S., & Palmese, G. R. (1999). Effects of temperature on cure kinetics and mechanical properties of vinyl–ester resins. *Journal of Polymer Science Part B: Polymer Physics*, *37*(7), 725-744.

[21] Gillham, J. K. (1986). Formation and properties of thermosetting and high Tg polymeric materials. *Polymer Engineering & Science*, *26* (20), 1429-1433.

[22] Wisanrakkit, G., & Gillham, J. K. (1990). The glass transition temperature (Tg) as an index of chemical conversion for a high-Tg amine/epoxy system: chemical and diffusion-controlled reaction kinetics. *Journal of Applied Polymer Science*, *41*(11-12), 2885-2929.

[23] Prime, R. B., Bair, H. E., Vyazovkin, S., Gallagher, P. K., & Riga, A. (2009). Thermogravimetric analysis (TGA). *Thermal analysis of polymers: Fundamentals and applications*, 241-317.

[24] Gan, S., Gillham, J. K., & Prime, R. B. (1989). A methodology for characterizing reactive coatings: Time–temperature–transformation (TTT) analysis of the competition between cure, evaporation, and thermal degradation for an epoxy-phenolic system. *Journal of applied polymer science*, *37*(3), 803-816.

[25] Pethrick, R. A., Hollins, E. A., McEwan, L., Pollock, A., Hayward, D., & Johncock, P. (1996). Effect of cure temperature on the structure and water absorption of epoxy/amine thermosets. *Polymer international*, *39* (4), 275-288.

[26] Tucker, S. J., Fu, B., Kar, S., Heinz, S., & Wiggins, J. S. (2010). Ambient cure POSS–epoxy matrices for marine composites. *Composites Part A: Applied Science and Manufacturing*, *41*(10), 1441-1446.

[27] Gupta, V. B., Drzal, L. T., Lee, C. C., & Rich, M. J. (1985). The temperature-dependence of some mechanical properties of a cured epoxy resin system. *Polymer Engineering & Science*, *25*(13), 812-823.

[28] Barton, J. M., Hamerton, I., Howlin, B. J., Jones, J. R., & Liu, S. (1998). Studies of cure schedule and final property relationships of a commercial epoxy resin using modified imidazole curing agents. *Polymer*, *39* (10), 1929-1937.

[29] Russo, C., Fernández-Francos, X., & De la Flor, S. (2019). Rheological and mechanical characterization of dual-curing thiol-acrylate-epoxy thermosets for advanced applications. *Polymers*, *11*(6), 997.

[30] Moller, J. C., Berry, R. J., & Foster, H. A. (2020). On the nature of epoxy resin post-curing. *Polymers*, *12* (2), 466.

[31] Zhang, J., Li, T., Wang, H., Liu, Y., & Yu, Y. (2014). Monitoring extent of curing and thermal–mechanical property study of printed circuit board substrates. *Microelectronics Reliability*, *54*(3), 619-628.

[32] Xie, M., Zhang, Z., Gu, Y., Li, M., & Su, Y. (2009). A new method to characterize the cure state of epoxy prepreg by dynamic mechanical analysis. *Thermochimica Acta*, *487*(1-2), 8-17.

[33] Rudolph, M., Naumann, C., & Stockmann, M. (2016). Degree of cure definition for an epoxy resin based on thermal diffusivity measurements. *Materials Today: Proceedings*, *3*(4), 1144-1149.

[34] Kister, G., & Dossi, E. (2015). Cure monitoring of CFRP composites by dynamic mechanical analyser. *Polymer Testing*, *47*, 71-78.

[35] Quintana, J. A., Boj, P. G., Villalvilla, J. M., Díaz-García, M. A., Ortiz, J., Martín-Gomis, L., ... & Sastre-Santos, Á. (2008). Determination of the glass transition temperature of photorefractive polymer composites from photoconductivity measurements. *Applied Physics Letters*, *92*(4).

[36] Yurov, V. M., Laurinas, V. C., Eremin, E. N., & Gyngazova, M. S. (2016). Determination of glass transition temperature for polymers by methods of thermoactivation spectroscopy. In *IOP Conference Series: Materials Science and Engineering* (Vol. 110, No. 1, p. 012018). IOP Publishing.

[37] Mahato, B., Lomov, S. V., & Abaimov, S. G. (2023, July). Quality Control and Cure Status Monitoring Sensor Based on Industrial Carbon Nanotube Masterbatch. In *2023 IEEE 23rd International Conference on Nanotechnology (NANO)* (pp. 261-265). IEEE.

[38] Mahato, B., Lomov, S. V., Jafarypouria, M., Owais, M., & Abaimov, S. G. (2023). Hierarchical toughening and self-diagnostic interleave for composite laminates manufactured from industrial carbon nanotube masterbatch. *Composites Science and Technology*, *243*, 110241.

[39] Zhao, Y., Liu, Q., Li, R., Lomov, S. V., Abaimov, S. G., Xiong, K., ... & Wu, Q. (2023). Self-sensing and self-healing smart fiber-reinforced thermoplastic composite embedded with CNT film. *Journal of Intelligent Material Systems and Structures*, *34*(13), 1561-1571.

[40] Mousavi, S. R., Estaji, S., Kiaei, H., Mansourian-Tabaei, M., Nouranian, S., Jafari, S. H., ... & Khonakdar, H. A. (2022). A review of electrical and thermal conductivities of epoxy resin systems reinforced with carbon nanotubes and graphene-based nanoparticles. *Polymer Testing*, *112*, 107645.

[41] Lebedev, O. V., Ozerin, A. N., & Abaimov, S. G. (2021). Multiscale numerical modeling for prediction of piezoresistive effect for polymer composites with a highly segregated structure. *Nanomaterials*, *11*(1), 162.

[42] Butt, H. A., Lomov, S. V., Akhatov, I. S., & Abaimov, S. G. (2021). Self-diagnostic carbon nanocomposites manufactured from industrial epoxy masterbatches. *Composite Structures*, *259*, 113244.

[43] Butt, H. A., Owais, M., Sulimov, A., Ostrizhiniy, D., Lomov, S. V., Akhatov, I. S., ... & Popov, Y. A. (2021, July). CNT/Epoxy-Masterbatch Based Nanocomposites: Thermal and Electrical Properties. In *2021 IEEE 21st International Conference on Nanotechnology (NANO)* (pp. 417-420). IEEE.

[44] Lomov, S. V., Gudkov, N. A., & Abaimov, S. G. (2022). Uncertainties in electric circuit analysis of anisotropic electrical conductivity and piezoresistivity of carbon nanotube nanocomposites. *Polymers*, *14*(22), 4794.

[45] Lomov, S. V., Akmanov, I. S., Liu, Q., Wu, Q., & Abaimov, S. G. (2023). Negative Temperature Coefficient of Resistance in Aligned CNT Networks: Influence of the Underlying Phenomena. *Polymers*, *15*(3), 678.

[46] Lomov, S. V., Akhatov, I. S., Lee, J., Wardle, B. L., & Abaimov, S. G. (2021, July). Non-linearity of electrical conductivity for aligned multi-walled carbon nanotube nanocomposites: Numerical estimation of significance of influencing factors. In *2021 IEEE 21st International Conference on Nanotechnology (NANO)* (pp. 378-381). IEEE.

[47] Zhang, L., Lu, Y., Lu, S., Wang, X., Ma, C., & Ma, K. (2022). In situ monitoring of sandwich structure in liquid composite molding process using multifunctional MXene/carbon nanotube sensors. *Polymer Composites*, *43*(4), 2252-2263.

[48] Luo, S., Obitayo, W., & Liu, T. (2014). SWCNT-thin-film-enabled fiber sensors for lifelong structural health monitoring of polymeric composites-From manufacturing to utilization to failure. *Carbon*, *76*, 321-329.

[49] Lee, J., & Wardle, B. L. (2019). Nanoengineered in situ cure status monitoring technique based on carbon nanotube network. In *AIAA Scitech 2019 Forum* (p. 1199).

[50] Lu, S., Chen, D., Wang, X., Xiong, X., Ma, K., Zhang, L., & Meng, Q. (2016). Monitoring the manufacturing process of glass fiber reinforced composites with carbon nanotube buckypaper sensor. *Polymer Testing*, *52*, 79-84.

[51] Lu, S., Zhao, C., Zhang, L., Ma, K., Bai, Y., Wang, X., & Du, K. (2019). In situ monitoring the manufacturing process of polymer composites with highly flexible and sensitive GNP/MWCNT film sensors. *Sensors and Actuators A: Physical*, *285*, 127-133.

[52] Xing, F., Li, M., Wang, S., Gu, Y., Zhang, W., & Wang, Y. (2022). Temperature Dependence of Electrical Resistance in Carbon Nanotube Composite Film during Curing Process. Nanomaterials 2022, 12, 3552.

[53] Lu, S., Zhao, C., Zhang, L., Chen, D., Chen, D., Wang, X., & Ma, K. (2018). Real time monitoring of the curing degree and the manufacturing process of fiber reinforced composites with a carbon nanotube buckypaper sensor. *RSC advances*, *8*(39), 22078-22085.

[54] Lu, S., Chen, D., Wang, X., Xiong, X., Ma, K., Zhang, L., & Meng, Q. (2017). Monitoring the glass transition temperature of polymeric composites with carbon nanotube buckypaper sensor. *Polymer Testing*, *57*, 12-16.

[55] Abaimov, S. G., & Abaimov, S. G. (2015). The theory of percolation. *Statistical Physics of Non-Thermal Phase Transitions: From Foundations to Applications*, 225-257.

[56] Gudkov, N. A., Lomov, S. V., Akhatov, I. S., & Abaimov, S. G. (2022). Conductive CNT-polymer nanocomposites digital twins for self-diagnostic structures: Sensitivity to CNT parameters. *Composite Structures*, *291*, 115617.

[57] Feng, C., & Jiang, L. (2013). Micromechanics modeling of the electrical conductivity of carbon nanotube (CNT)–polymer nanocomposites. *Composites Part A: Applied Science and Manufacturing*, *47*, 143-149.

[58] Zhang, S., Park, J. G., Nguyen, N., Jolowsky, C., Hao, A., & Liang, R. (2017). Ultra-high conductivity and metallic conduction mechanism of scale-up continuous carbon nanotube sheets by mechanical stretching and stable chemical doping. *Carbon*, *125*, 649-658.

[59] Wang, L., Peng, W., Sarafbidabad, M., & Zare, Y. (2019). Explanation of main tunneling mechanism in electrical conductivity of polymer/carbon nanotubes nanocomposites by interphase percolation. *Polymer Bulletin*, *76*, 5717-5731.

[60] Uçar, E., Dogu, M., Demirhan, E., & Krause, B. (2023). PMMA/SWCNT Composites with Very Low Electrical Percolation Threshold by Direct Incorporation and Masterbatch Dilution and Characterization of Electrical and Thermoelectrical Properties. *Nanomaterials*, *13*(8), 1431.

[61] Haghgoo, M., Ansari, R., Hassanzadeh-Aghdam, M. K., & Nankali, M. (2019). Analytical formulation for electrical conductivity and percolation threshold of epoxy multiscale nanocomposites reinforced with chopped carbon fibers and wavy carbon nanotubes considering tunneling resistivity. *Composites Part A: Applied Science and Manufacturing*, *126*, 105616.

[62] Lee, J., & Wardle, B. L. (2019). Nanoengineered in situ cure status monitoring technique based on carbon nanotube network. In *AIAA Scitech 2019 Forum* (p. 1199).

[63] Burstein, E., & Lundqvist, S. (Eds.). (1969). *Tunneling phenomena in solids* (pp. 427-442). New York: Plenum Press.

[64] Terrones, J., Windle, A. H., & Elliott, J. A. (2014). The electro-structural behaviour of yarn-like carbon nanotube fibres immersed in organic liquids. *Science and Technology of Advanced Materials*, *15*(5), 055008.